# Optically-generated Focused Ultrasound for Noninvasive Brain Stimulation with Ultrahigh Precision

**Authors**

Yueming Li[1, †], Ying Jiang[2, †, ‡], Lu Lan[3, †], Xiaowei Ge[3], Ran Cheng[4], Yuewei Zhan[5], Guo Chen[3], Linli Shi[4], Runyu Wang[3], Nan Zheng[6], Chen Yang[3,4, *], Ji-Xin Cheng[3,5, *]

**Affiliations**

[1] Department of Mechanical Engineering, Boston University, Boston, MA 02215, USA.

[2] Graduate Program for Neuroscience, Boston University, Boston, MA 02215, USA.

[3] Department of Electrical and Computer Engineering, Boston University, Boston, MA 02215, USA.

[4] Department of Chemistry, Boston University, Boston, MA 02215, USA.

[5] Department of Biomedical Engineering, Boston University, Boston, MA 02215, USA.

[6] Division of Materials Science and Engineering, Boston University, Boston, MA 02215, USA.

† These authors contributed equally to this work.

‡ Current affiliation: Department of Biological Engineering, Massachusetts Institute of Technology, Cambridge, MA 02142, USA

* Corresponding author. Email: jxcheng@bu.edu (J.C.); cheyang@bu.edu (C.Y.).

**Abstract**

High precision neuromodulation is a powerful tool to decipher neurocircuits and treat neurological diseases. Current non-invasive neuromodulation methods offer limited precision at the millimeter-level. Here, we report optically-generated focused ultrasound (OFUS) for non-invasive brain stimulation with ultrahigh precision. OFUS is generated by a soft optoacoustic pad (SOAP) fabricated through embedding candle soot nanoparticles in a curved polydimethylsiloxane film. SOAP generates a transcranial ultrasound focus at 15 MHz with an ultrahigh lateral resolution of 83 μm, which is two orders of magnitude smaller than that of conventional transcranial focused ultrasound (tFUS). Here, we show effective OFUS neurostimulation *in vitro* with a single ultrasound cycle. We demonstrate submillimeter transcranial stimulation of the mouse motor cortex *in vivo*. An acoustic energy of 0.6 mJ/cm$^2$, four orders of magnitude less than that of tFUS, is sufficient for successful OFUS neurostimulation. OFUS offers new capabilities for neuroscience studies and disease treatments by delivering a focus with ultrahigh precision non-invasively.

**Introduction**

To understand how the brain functions and how its dysfunction causes diseases, modalities to modulate neuronal activity with ultrahigh precision are needed. Brain stimulation modalities with millimeter precision usually activate multiple functional regions and cause unintended responses [1]. For example, limited spatial precision prevents the activation of individual brain regions, making it difficult to map the motor cortex in mice [2]. A spatial precision better than 0.2 mm is desired for precise targeting in stimulating specific cell layers in the rat hippocampus [3]. For decoding the human brain from an engineering point of view, an ultrahigh spatial resolution of stimulation to match the resolution for reading (0.1 mm) is desirable for single neuron-based systems in brain-computer interfaces [4]. Therefore, a neuromodulation tool with ultrahigh precision is needed for mapping the brain sub-regions by modulating a small population of neurons. Electrical stimulation tools are a gold standard for neuromodulation studies and disease treatments. Deep brain stimulation with implanted electrodes has been approved for clinical treatment of Parkinson's disease, depression, and epilepsy [5, 6]. However, the current leakage over several millimeters limits the precise control of targeting in electrical stimulation [7]. Optogenetics provides an unrivaled sub-cellular spatial resolution and specificity in targeted cell types, which has advanced the study of neuroscience [8]. Recently developed transcranial optogenetics in mice can target a brain area of 0.8 to 1 mm laterally at a penetration depth of 5 ~ 6 mm without surgery [9]. However, transcranial optogenetics only provides a light transmission rate of ~ 0.02% at 7 mm. Therefore, there is an increased risk of heat accumulation along the light path in the illuminated tissue at powers that deliver sufficient light energy. Furthermore, both conventional and transcranial optogenetics rely on viral transfection, which has yet limited their applications in the human brain.

Non-invasive non-genetic neuromodulation is attractive as it avoids the risk of surgery and is applicable to the human brain. Transcranial direct current stimulation (tDCS) and transcranial magnetic stimulation (TMS) [10] provide a spatial resolution at the centimeter-level due to the long wavelength of the electromagnetic waves used. The emerging transcranial focused ultrasound (tFUS) as a non-invasive neuromodulation method offers millimeter-level precision in various models, such as mice [11, 12], rats [13], rabbits [14], monkeys [15], and even humans [16, 17]. A low ultrasonic frequency of ~ 1 MHz or less is preferred in tFUS for high transcranial efficiency, but limits its spatial resolution at 1 to 5 millimeters [18]. Ultrasound with higher frequencies also draws attention for its high spatial resolution. 5 MHz ultrasound has been demonstrated to induce electromyography (EMG) responses in mouse brains with a submillimeter spatial resolution of 0.3 mm [19]. Stimulation in the retina was demonstrated ex vivo with a much higher acoustic frequency of 43 MHz [20]. Very recently, Cheng et al used a 30 MHz transducer to induce neural inhibition effects, whereas ultrasound was delivered through a cranial window [21]. Thus far, non-invasive brain stimulation by focused ultrasound with a spatial resolution of 0.1 mm has not been demonstrated. Non-invasive non-genetic neuromodulation with ultrahigh precision remains a critical unmet need.

The optoacoustic effect is an alternative way to generate ultrasound. Optoacoustic materials absorb a short pulse of light and convert it into a transient temperature increase and thermal expansion and compression, resulting in the generation of an ultrasound pulse [22]. To maximize the light-to-sound conversion efficiency, photoacoustic materials composed of light-absorbing units and thermal expansion units are extensively explored. Polydimethylsiloxane (PDMS) has drawn much attention as a thermal expansion material because of its high thermal expansion and transparency. For light absorbers, materials of interest include metal and carbon materials, including gold, titanium, chrome [23], carbon black, carbon nanotube [24], carbon nanofiber, carbon nanoparticles, and candle soot [25]. While metal-based materials have a high light-absorption at certain wavelengths via the resonance effect, carbon-based materials provide a broadband absorption and compatibility with a variety of laser systems. Recently Jiang et al. and Shi et al. reported fiber-based optoacoustic emitters coated with a mixture of carbon-based materials and epoxy/PDMS for submillimeter and single neuron stimulation [26-28]. The design of tapered fiber optoacoustic emitters with a diameter of 20 μm, enables selective activation of subcellular structures as a point source of ultrasound [28]. However, since they exploit near-field ultrasound for localized neuromodulation, fiber optoacoustic emitters need to be surgically implanted to the target and cannot be applied transcranially.

Here, we report the development of optically-generated focused ultrasound (OFUS) for non-invasive neuromodulation with ultrahigh precision below 0.1 mm for the first time. OFUS is generated by a curved soft optoacoustic pad (SOAP) upon a nanosecond laser excitation. SOAP is fabricated using PDMS and a carbon-based absorber. The transverse diameter was tailored to provide a numerical aperture (NA) of 0.95 to enable a tighter spatial focusing and to maximize the focal pressure. This NA is close to the theoretical limit of 1 and much larger than that of a typical conventional lead zirconate titanate (PZT)-based transducer. To identify the optimal absorber for efficient conversion of photons to acoustic waves, SOAPs based on four different optoacoustic materials were fabricated and tested, including heat shrink membrane (HSM), carbon nanotubes mixed with PDMS (CNT-PDMS), carbon nanoparticles mixed with PDMS (CNP-PDMS), and candle soot layered with PDMS (CS-PDMS). Their optoacoustic conversion efficiencies were compared by measuring the pressure at the foci. The CS-PDMS SOAP was found to be the most efficient, generating ~ 48 MPa at the ultrasound focus under a 0.62 mJ/cm$^2$ laser input. We further demonstrated that CS-PDMS OFUS produces an ultrahigh spatial resolution of ~ 83 μm with transcranial capability, which is a two orders of magnitude improvement from the resolution of a few millimeters offered by tFUS. We achieved direct and transcranial single-cycle OFUS stimulation reliably and safely, verified by calcium imaging in cultured neurons *in vitro*. The total ultrasound energy input of OFUS is found to be four orders of magnitude less than that of tFUS when evoking a similar level of neural response [29]. We also demonstrated non-invasive transcranial neurostimulation with OFUS in mice. We confirmed a stimulation volume with a width of 200 μm in the mouse brain with immunofluorescence imaging. Lastly, we validated the functional outcomes of OFUS stimulation in the mouse motor cortex by electrophysiological recording *in vivo*.

## Results

### Fabrication of SOAP and optimization of optoacoustic efficiency

**Figure 1A** shows a schematic of SOAP, composed of a PDMS substrate and a curved layer of optical absorbers. Ultrasound waves are generated at the surface of the curved layer when a nanosecond laser is delivered. The waves emitted from different angles on the curvature arrive at its geometric center coherently. A working distance of 2 mm is designed to ensure that the generated ultrasound can penetrate the mouse skull and reach the cortical layer.

To optimize the geometric design of SOAP, we firstly used numerical simulations to predict generated acoustic fields. Two-dimensional numerical simulations were performed using the k-wave toolbox in MATLAB. A single layer of grid on the curvature was assigned as the source of ultrasound to model the optoacoustic layer. The generated ultrasound propagates into the water with air-backing to mimic the application scenarios. A central frequency of 15 MHz was set according to reported photoacoustic frequencies generated by carbon-based absorbers [30]. A working distance of 2 mm between the ultrasound focus and the top flat surface of SOAP was fixed. We then tuned the radius and the transverse diameter of SOAP to obtain the lateral and axial resolution of the ultrasound field at the focus, defined as the full width at half maximum (FWHM) of the field.

The simulated lateral resolution *R* and axial resolution as a function of the NA, calculated based on the ratio of the transverse diameter to the radius, is plotted in **Figure 1B** and **Figure**

**1C**. A larger NA can provide a better lateral and axial resolution. The lateral resolution is reversely proportional to the NA, shown in a fitting curve (red) of $R = 71.51/NA$ ($R^2 = 0.9899$, fitting coefficient of determination). This relationship agrees with the equation of the lateral resolution in acoustic-resolution photoacoustic microscopy. The lateral resolution depending on the NA follows the equation below[22].

$$R_L = 0.71 \frac{v}{NA \cdot f} \quad (1)$$

Here, $v$ is the ambient sound speed and $f$ is the central frequency. The orange area in **Figure 1B** indicates the range of the NA in conventional PZT-based transducers [20, 21, 31-33]. It is difficult for the single crystal piezoelectric material to reach a high NA due to the cracking in PZT-based single-element focused ultrasound transducers [34]. However, a high NA up to 0.95, close to the theoretical limit value of 1, can be obtained easily by soft optoacoustic materials. While a higher NA enables a tighter acoustic focus, we chose to demonstrate a NA of 0.95 considering a minimal stand-off distance of 2 mm is needed for our device to target the cortical layer of the mouse brain and the size of the mouse head (slightly wider than 1 cm). OFUS is expected to deliver a lateral resolution of 75 µm corresponding to a NA of 0.95. In comparison, a conventional transducer at the same ultrasonic frequency delivers a lateral resolution of 132 µm, twice of the resolution of OFUS.

To simulate the ultrasound field generated by SOAP with a high NA of 0.95, the radius of 6.35 mm and the transverse diameter of 12.1 mm was selected. The result confirms that this geometry provides a focused ultrasound field at the geometric center as expected (**Figure S1**). The simulated OFUS at the focus has a lateral resolution of 78 µm, which is consistent with the calculation above. An axial resolution of 209 µm was obtained. This resolution provides an ultrahigh precision for neurostimulation in small animals [35].

Such a high NA provides OFUS not only with a high lateral resolution but also a high focal gain $G$ at the focus. $G$ is defined by the ratio of the pressure at the focal point to the pressure on the spherical surface. A spherical surface with a high NA corresponding to a low $f$-number, defined as the ratio of the radius to the transverse diameter, provides a high focal gain $G$ according to the equation below [24].

$$G = \frac{2\pi f}{c_0} r \left(1 - \sqrt{1 - \frac{1}{4 f_N^2}}\right) \quad (2)$$

In this equation, $f, c_0, r,$ and $f_N$ stands for the acoustic frequency, the speed of sound in the medium, the radius of curvature, and the $f$-number, respectively. We estimate the theoretical maximum focal gain $G_{max} \approx 280$ for SOAP assuming uniform laser illumination with a frequency, a $f$-number, a water attenuation coefficient, and a working distance of 15 MHz, 0.52, $2.2 \times 10^{-3}$ dB/(cm × MHz$^2$) and 2 mm, respectively. This theoretical maximum focal gain is 5- and 92-fold higher than that of a conventional PZT-based transducer with $f_N$ of 1 and 4, respectively [20, 21, 31-33]. In summary, OFUS taking advantage of a high NA close to the limit enables a lateral resolution less than 100 µm at 15 MHz and a focal theoretical maximum pressure up to 92-fold higher compared to a PZT-based transducer at the same frequency.

In addition to the geometry, we also evaluated 4 different absorbers to optimize the optoacoustic conversion efficiency experimentally. We fabricated SOAP of the same NA with HSM, CNT-PDMS, CNP-PDMS, and CS-PDMS (**Figure 1D**). For HSM, the elastic black polyolefin itself serves as the light-absorber and expansion material simultaneously. In the other three designs, the carbon-based materials are embedded in PDMS, serving as light-absorbers and expansion material, respectively. **Figure S2** shows the fabrication method for CS-PDMS SOAP. Details of fabrication methods for SOAPs are described in the Methods.

To investigate whether the transfer process produced a well-mixed matrix of CS and PDMS, we sliced the CS-PDMS SOAP into layers with a thickness of 200 µm and examined the morphology using scanning electron microscopy (SEM). A mixed layer composed of PDMS and evenly embedded CS particles can be distinguished from the smooth pure PDMS layer (**Figure 1E**). The mixture layer has a thickness of 2.7 µm, which is very close to the theoretical thickness of 2.15 µm for the optimal optoacoustic transduction of CS-PDMS [25]. We identified that the diameter of CS nanoparticles was 55 nm (**Figure S3)** and estimated a deposition rate of 200 µm/s, which are consistent with the reported size and rate prepared by the same method, respectively [36].

Besides the morphology, the chemical composition of the CS-PDMS composite was examined by label-free stimulated Raman scattering (SRS) and photothermal imaging to confirm the boundary between the CS-PDMS layer and the pure PDMS layer (**Figure S4**). The femtosecond SRS of C-H bonds in PDMS only occurs when the two beams, a pump and a Stokes laser beam, temporally overlap (t ~ 0 s). The candle soot's photothermal signal has a much slower decay, which is close to constant with the delay between two laser beams. Therefore, we can chemically map the spatial distribution of the CS particles in the PDMS by SRS and photothermal imaging of the same sample that we took SEM images of. In **Figure 1F**, the merged image illustrates a mixed CS-PDMS layer with a thickness ~ 3 µm and a pure PDMS layer. This image provides chemical information to confirm the boundary between those layers, which is consistent with the previous SEM result (**Figure 1E**).

To characterize the optoacoustic efficiency of the four fabricated SOAPs, we delivered laser pulses at 0.62 mJ/cm$^2$ to each design. A needle hydrophone was used to record the generated waveforms (**Figure 1G**). HSM provided the smallest amplitude, 5 times smaller compared to CNT- and CNP-PDMS. Signals from CNT- and CNP-PDMS were at the same level. CS-PDMS generated 48 MPa, which is 6-fold larger than that of CNT- and CNP-PDMS. This result is consistent with previous reports [36]. The ultrasound pulse widths of HSM, CNT-, CNP-, and CS-PDMS were 0.31 µs, 0.24 µs, 0.29 µs, and 0.09 µs, respectively. The frequency spectrum of the OFUS signals is shown in **Figure 1H**. HSM produced a central frequency at ~ 3 MHz, and the CNT- and CNP-PDMS provided higher central frequencies at ~ 5 MHz. The CS-PDMS generated the highest central frequency at ~ 15 MHz. – 6 dB widths were found at 5 and 35 MHz and the ultrasound signal of CS-PDMS had a broad bandwidth of 200%. We selected CS-PDMS for further experiments because it had the highest optoacoustic conversion efficiency and highest central frequency, which provides a tight focus. The CS-PDMS SOAP has a transverse diameter of 12.2 mm, corresponding to a high NA of 0.96. We further evaluated the light leakage of CS-PDMS with a power meter. Only 2% of light energy was detected behind CS-PDMS, demonstrating that up to 98% of the light energy was absorbed by CS-PDMS. A linear relation between the input laser pulse energy and CS-PDMS generated OFUS pressure at the focus was also confirmed (**Figure S5**). In conclusion, we have fabricated a SOAP made of CS-PDMS with a high NA, high focal gain, and high optoacoustic conversion efficiency for neural stimulation.

**OFUS demonstrates an ultrahigh spatial resolution and high transcranial efficiency**

To confirm that OFUS provides an ultrahigh spatial resolution after penetrating the skull for non-invasive applications, we characterized the resolution of OFUS before and after penetrating a piece of mouse skull. We placed a piece of mouse skull (thickness ~ 0.15 mm) between the hydrophone and SOAP (**Figure 2A**). The transcranial efficiency of OFUS was evaluated by measuring the amplitude of the transcranial ultrasound signal and normalizing it to the peak amplitude of the signal at the focus in the absence of the skull (**Figure 2B**). The transcranial efficiency of OFUS was 69 %. Such a high efficiency is sufficient for later non-invasive applications. The frequency spectrum shows that a high central frequency remains after the penetration of the mouse skull, which ensures an ultrahigh transcranial precision (**Figure S6**).

To characterize the focus size without and with a mouse skull in the path, we swept the focus to acquire lateral and axial profiles (**Figure 2C** and **2D**). All profiles were normalized to the peak amplitude of the signal without the skull for comparison. We define lateral and axial resolutions according to the FWHM in the respective directions. The lateral and axial resolutions of OFUS without the skull were found to be 66 μm and 284 μm, close to simulation data. With the skull, the FWHMs were 83 μm and 287 μm, respectively, showing no substantial change in the focus size after penetrating through the mouse skull. These resolutions are two orders of magnitude higher than the resolution of the low-frequency ultrasound typically used in neuromodulation, which is ~ 5 mm and ~ 40 mm for a 0.5 MHz transducer, respectively [18].

To examine the location of the OFUS focus, we visualized the propagation of ultrasound focus by an optoacoustic tomography system. Both the ultrasound image and optoacoustic (OA) image were detected with a 128-element ultrasound transducer array (**Figure S7a**). All the images acquired with the transducer array only shows the middle part of the curvature of SOAP and the wavefront due to the limited acceptance angle of the array. **Figure S7b** and **S7c** shows merged ultrasound images of SOAP and the mouse skull (Green) and OA image of OFUS (Red) without and with the skull. No significant change in the position of OFUS focus was observed. In **Figure S7d** and **S7e**, we reconstructed the propagation process of the OFUS from SOAP's surface to the focus area. The delays between OA images are not evenly distributed (Supplementary **Video 1**). These propagation results not only illustrate the interference of the optoacoustic signal to generate a tight focus, but also demonstrate that this interference would not be influenced significantly by the presence of the mouse skull, therefore enabling a tight transcranial focus.

**SOAP enables direct and transcranial stimulation of primary cortical neurons**

With the goal of OFUS neurostimulation, we investigated whether OFUS can evoke responses in cultured neurons. SOAP was placed ~ 2 mm above primary cortical neurons transfected to express GCaMP6f (**Figure 3A**). The calcium signal of neurons was recorded by a fluorescence imaging system. To locate the ultrasound focus, we visualized it via fluorescent beads, which were pushed away by acoustic radiation force (Supplementary **Video 2**). Based on the movement of the beads, we identified the diameter of the focus to be ~ 100 μm, which is consistent with our tested lateral resolution.

For neural stimulation, we delivered a single laser pulse to acquire a single optoacoustic cycle. We increased the laser pulse energy from delivering a peak to peak pressure of 23.8 MPa, with a non-linear step determined by filters, until a successful activation defined as max *ΔF/F* > 10% was achieved. **Figure 3B** shows representative calcium images before and after the OFUS stimulation with a pressure of 29.8 MPa. Neural activation was only observed at the center of the field of view corresponding to the focus location, showing a localized stimulation ability of OFUS with ultrahigh precision (**Figure 3B** and Supplementary **Video 3**). Light leakage is shown at the delivery of laser in Supplementary **Video 3**. To assure the activation is not evoked by this leakage, we tuned the laser energy to 2% and delivered it to the neurons directly. No activation of neurons was observed in this control experiment. We calculated the success rate of OFUS stimulation by comparing the number of the stimulated neurons to all the neurons within the focus area in each trial. This focus area is defined as a circle with a diameter of 66 μm. CS-PDMS OFUS achieved a success rate of 71.4% at a pressure of 29.8 MPa within 6 trials.

Calcium traces from 37 neurons that were successfully activated at the OFUS focus were analyzed ($N = 7$ dishes). Two types of response were observed, a transient response and a prolonged response. We fit the decay of the response curves exponentially and obtained the *1/e* time constant. Transient activations show decay time constants ranging from 2 to 5 s, while prolonged activations have time constants of 5 s and up [37]. Transient responses show an average decay time constant of 4.2 s and a max *ΔF/F* of 31% ± 8% ($N = 6$ from 3 cultures, data in mean ± SD) (**Figure 3C**). Prolonged responses have an average time constant of 7.8 s and max *ΔF/F* of 62% ± 12% ($N = 31$ from 4 cultures) (**Figure 3D**). Both activations were observed at the focus right after a single cycle of OFUS was delivered. This action potential of prolonged activation then propagates through the network (**Figure S8**).

We also examined the threshold pressure to evoke transient or prolonged activation (**Figure 3E**). The threshold for transient responses is 32.3 ± 4.3 MPa, significantly lower than that of prolonged responses, which is 48.5 ± 15.3 MPa, (two-sample *t*-test, $N = 37$ from 7 cultures, \*\*\*p<0.001). The ability of neurons to differentiate the magnitude of mechanical stimuli and respond to higher amplitude stimulation with a slower decay agrees well with several reports studying neural response to mechanical stimuli [37, 38].

To demonstrate the safety of OFUS neural stimulation, we studied whether neurons can be stimulated repeatedly. We delivered single-cycle stimulation to the same group of neurons 3 times with an interval of 2 min at a pressure of 29.8 MPa. **Figure 3F** shows max *ΔF/F* images of each stimulation taken from the calcium imaging. No visible change in morphology was observed. The calcium trace (**Figure 3G**) showed a similar amplitude of max *ΔF/F* after each stimulation, which confirms no functional damage after repeated stimulation. We further tested the viability of neurons after repeated stimulation at 55.3 MPa and 71.5 MPa, respectively. These two energy levels were selected to be 50 ~ 70% higher than the thresholds to leave an extra margin for the safety demonstration. We studied five groups of neurons at each energy level. For each group, we delivered 30 cycles in total. To calculate the cell viability, we counted live and dead cells after 30 min incubation. A group without laser excitation was performed as a control. No significant difference in the cell viability was observed between the stimulated groups and the control group for both energy levels (**Figure S9**). These data collectively show that OFUS can stimulate neurons repeatedly and reliably, without any damage to the morphology or functionality of neurons.

To compare OFUS to transducer-based FUS, we performed neural stimulation with a 20 MHz conventional focused ultrasound transducer in exactly the same experimental setup. We started with a low ultrasound intensity and a short duration of continuous wave (CW) ultrasound. Then we increased them step by step until a neural response with max *ΔF/F* > 10% was recorded. An ultrasound intensity of 32 W/cm$^2$ with a 500 ms duration evoked a calcium response up to 20% in our experimental setup (**Figure S10**). Compared to a conventional transducer, OFUS evokes neural response at a similar level, but with four orders of magnitude less acoustic energy density input. In addition, compared to a study with a conventional transducer to evoke ~ 15% calcium response [29], OFUS stimulation also used four orders of magnitude less energy density (**Table 1**). Notably, this earlier work used cortical neurons extracted from mouse brains and cultured on a film dish, while our work used cortical neurons from rat brains and cultured on a glass-bottom dish. Together, these results demonstrate the high stimulation efficiency of OFUS with a unique single-cycle stimulation mode that benefits from the optoacoustic effect.

To test the transcranial stimulation capability, we studied the stimulation threshold of cultured neurons with OFUS penetrating a piece of mouse skull. We embedded a piece of mouse skull in the SOAP and kept the same setup as previously described (**Figure 4A**). **Figure 4B** and Supplementary **Video 4** show representative images and a video of calcium signal before and after transcranial stimulation by OFUS, and the corresponding max $\Delta F/F$ image. Only the neuron in the middle of the field of view was activated, confirming that the transcranial focus is still tight for ultrahigh-precision stimulation. With a single cycle stimulation by OFUS, successful stimulations with an average max $\Delta F/F$ of 27% ± 5% were recorded ($N$ = 18 from 7 cultures, **Figure 4C**). The thresholds for direct stimulation and transcranial stimulation were compared in **Figure 4D**. While the average threshold for direct stimulations was 46.0 ± 12.8 MPa, the transcranial stimulation had a threshold of 65.5 ± 9.4 MPa (two-sample $t$-test, $N$ = 46 from 11 cultures, \*\*\*p<0.001). This increase of the laser energy in transcranial stimulation comes naturally due to the energy loss in the transcranial process. Assuming the neurostimulation threshold remains the same during different trials, the pressure of 46.0 MPa needs to reach 65.5 MPa to compensate for the previously tested transcranial efficiency of 69%. This theoretical value is consistent with our experimental transcranial stimulation threshold. Collectively, our results demonstrate the ability of OFUS to stimulate cortical neurons both directly and transcranially.

**OFUS mediates high-precision neural stimulation *in vivo***

With successful stimulation of cultured neurons directly and transcranially, we further asked whether OFUS can activate neurons in the brains of living animals. Adult C57BL/6J mice were used for stimulation *in vivo*, and the effect of stimulation was evaluated by both immunofluorescence imaging and electrophysiology recording. The focus of the generated ultrasound was carefully aligned with the motor cortex based on stereotaxic coordinates (Medial-Lateral: 1.5, Anterior-Posterior: 0.5) (**Figure 5A** and **Figure S11**).

We first visualized the stimulated area by labeling c-Fos proteins, which has been widely used to identify stimulated neurons. To induce a robust c-Fos expression, we applied OFUS stimulation with a pressure of 8.5 MPa for a pulse train of 20 cycles, which lasted 30 min with a 33% duty cycle (**Figure S12**). The mice were put to rest for 1 h to maximize c-Fos expression. Brain slices were stained with c-Fos and DAPI. DAPI can label all the nuclei of neurons and provide a reference. Thus, the c-Fos positive neurons were counted only when it co-localized with DAPI signal. Many more c-Fos positive cells were observed at the OFUS-stimulated area compared to the control group at the contralateral area (**Figure 5B**). The percentage of c-Fos positive cells in the OFUS group was 34 ± 4 %, a significant increase from 2 ± 0.3 % in the control (**Figure 5C**, two-sample $t$-test, n = 3, \*\*\*p<0.001). Importantly, the c-Fos signal was confined to the target site with an area of ~ 200 μm in diameter. This result demonstrates a superior spatial resolution compared to the conventional PZT-transducer-based tFUS stimulation (1 ~ 5 mm) [18]. No significant c-Fos expression outside the targeted area was observed, confirming direct OFUS stimulation. These results collectively suggest the ability of OFUS to directly evoke responses of neurons non-invasively with a high spatial resolution of 200 μm.

Further evaluation of the functional outcome by OFUS stimulation was conducted using EMG. The focus of OFUS was aligned to the primary motor cortex of the mouse brain to evoke cramps of muscles. The EMG electrode was inserted parallel to the biceps femoris muscle into the hind limb, and the ground electrode was inserted into the tail (**Figure 5A**). A laser pulse train with a duration of 2 s at a pressure of 8.5 MPa was delivered to SOAP. Strong EMG responses (0.458 ± 0.03 mV) were recorded from the contralateral hind limb (**Figure 5D**). These EMG signals have a typical delay of ~ 61 ± 6 ms between laser onsets and EMG response. After processing with a bandpass filter and full-wave rectifier, the envelopes of the EMG signals were plotted (**Figure 5E**). To eliminate the possibility of EMG response being evoked by the auditory effect of ultrasound, we stimulated at the somatosensory cortex based on stereotaxic coordinates as a control. No significant EMG response is observed in the control group. This result suggests that EMG responses were evoked directly by OFUS stimulation without the involvement of the auditory pathway.

Next, we evaluated the safety of OFUS stimulation *in vivo* by hematoxylin and eosin (H&E) staining. After the EMG recording, brains were extracted and fixed. Brain slices of 5 μm thickness were obtained every 150 μm and standard H&E staining was performed. We examined all the brain slices and compared the targeted area to the control group at the contralateral area. No significant change in morphology of cells between these groups was observed (**Figure 5F**). This result illustrates that OFUS stimulation does not induce visible damage to mouse brains.

To further evaluate the biosafety in terms of cavitation and thermal accumulation, we calculated the mechanical index (MI) and tested the temperature rise. With our laser input used in the *in vivo* experiment, the estimated transcranial peak-to-peak pressure 6.0 MPa delivered to the mouse brain was below the level of 40 MPa, at which no tissue lesion was reported [39]. The transcranial peak negative pressure of OFUS was estimated to be 2.7 MPa, which is below the threshold of bubble cloud generation in soft tissue (25-30 MPa) [40]. An MI of 0.5 is obtained for OFUS stimulation based on acoustic attenuation coefficient for brain tissue of 0.91 dB/(cm × MHz) . This MI is well below the FDA approved value of 1.9 and is considered unlikely to damage tissue by inducing cavitation.

For thermal safety, the temperature profiles were recorded with a thermocouple. We applied the pressure of 52.7 MPa, which matched the highest energy threshold used in this work. We delivered OFUS for 10 s. A temperature elevation up to 0.4 K was observed at the surface of SOAP (**Figure S13**). No temperature increase was observed at the focus of OFUS. Therefore, even at the longest duration of 2 s used in the successful stimulation, temperature rise induced by OFUS is expected to be less than 0.1 K at the focus of OFUS. Therefore, OFUS has been demonstrated to be safe for brain modulation both biologically and physically.

**Discussion**

In this work, we developed a CS-PDMS SOAP for OFUS generation, characterized its spatial resolution and transcranial ability, and validated the transcranial neural stimulation *in vitro* and *in vivo*. This is the first time that successful non-invasive brain stimulation with ultrahigh precision below 0.1mm is achieved. The large NA from the SOAP allows a tight transcranial lateral focus of 83 μm, which is beyond the reach of the piezo-based low-frequency tFUS. OFUS provides a transcranial efficiency of 69%, which enables transcranial applications.

Direct and transcranial stimulation of cortical neurons *in vitro* was recorded. A success rate of 71% was achieved with a pressure of 29.8 MPa. The ultrasound energy input of OFUS needed to evoke neural response is four orders of magnitudes lower than that of conventional ultrasound. Successful non-invasive OFUS stimulation at the mouse motor cortex *in vivo* was demonstrated by immunofluorescence imaging and EMG recording. The spatial resolution *in vivo* was found to be 200 μm. The safety of OFUS stimulation was confirmed by cell viability *in vitro* and histology analysis *in vivo*, and by MI and temperature rise measurements.

An important observation about OFUS stimulation is that the stimulation was evoked by the direct effect of the acoustic wave instead of a confounding auditory effect [41, 42]. In experiments *in vitro*, cultured neurons responded to the OFUS stimulation without auditory circuits. In studies *in vivo*, c-Fos positive neurons were located at the stimulation site corresponding to the direct stimulation. Moreover, no EMG response was recorded in the control group, which was stimulated at the somatosensory cortex, indicating that bone conduction to the cochlea is not involved in the process. Such an observation agrees with reported direct stimulation with ultrasound[26, 29].

The mechanism of evoking a neural response with ultrasound is still under investigation. While focused ultrasound can induce several physical effects on biological tissue, such as thermal accumulation, cavitation, and mechanical force, it has been debated whether these bioeffects could lead to neural stimulation [29, 43, 44]. The thermal effect has been reported as a plausible mechanism of ultrasound neuromodulation [45]. The maximum temperature rise in OFUS was evaluated to be < 0.1 K. This is far below the threshold required to thermally modulate neuron activities ($\Delta T \geq 5$ K) [44]. Therefore, the thermal stimulation mechanism for OFUS can be ruled out. The intramembrane cavitation is a prevailing explanation of how ultrasound perturbs neurons. A model was set up to study how mechanical energy of CW ultrasound is absorbed by the cellular membrane and induces intramembrane cavitation [43]. It requires 2 to 12 cycles for cavitation to reach a stable level and further oscillate with US. The cavitation mechanism induced by consecutive cycles does not apply to OFUS, because the single cycle stimulation of OFUS would be too short to form a stable gas bubble and keep it oscillating to perturb the ion channels. The radiation force is another plausible explanation believed to dominate high-frequency ultrasound stimulation [20, 46]. Successful stimulation of neurons has been demonstrated with radiation force by activating mechanosensitive channels [20, 29] or changing the capacitance in the lipid bilayers[47]. The effect of the radiation force of OFUS has already been visualized by the movement of beads in our work. Further work to study whether the radiation force delivered by OFUS can activate mechanosensitive channels or voltage-gated channels will elucidate the mechanism of OFUS stimulation.

Brains are complex and further developments in modulation and therapy require multisite stimulation. Patterned neuromodulation, for example, can further provide selectivity in motor control for therapy [48]. For conventional PZT-based ultrasound massive arrays, the massive cables connected to each element impede the application of a wearable clinical device. Taking advantage of optical engineering, the OFUS device can be scaled up into a massive array for multisite neuromodulation. A light-weight OFUS device also provides improved accessibility and wearability for long-term treatments. In addition, OFUS devices, with no metal components, further offer improved compatibility with real-time magnetic resonance imaging (MRI) guidance and functional MRI monitoring. These features of OFUS enable real-time fMRI evaluation of stimulation treatment and provide opportunities for close-loop treatments in clinical applications.

Notably, OFUS offers an opportunity to improve spatiotemporal control in histotripsy. Histotripsy uses transducer to deliver low frequency (< 3 MHz), short pulses (< 10 cycles) of high intensity ultrasound (> 20 MPa) for cavitation-based therapy to remove tissue [49]. Those transducers are driven under thousands of volts, suffering from the risk of dielectric breakdown. Our OFUS generated by SOAP can deliver the ultrahigh spatial control with a higher frequency, improve the temporal precision with single cycle, and provide ultrasound with high intensity by simply increasing the energy of input light without the risk of dielectric breakdown [24]. This niche highlights a future direction of OFUS application in ultrasound surgery with improved spatiotemporal control, minimized damage, and heat accumulation to surrounding tissues.

Notably, a larger NA of SOAP provides a tighter acoustic focus without further increasing the frequency, but makes it more vulnerable to aberrations in the skull. The application of SOAP is limited by the distortion in subjects with thicker skulls, especially in humans. Yet, by engineering the geometric surface, our SOAP can emit any desired acoustic wavefronts. This flexibility guarantees a pre-compensated optimization to minimize the aberration. Thus, in future work, one can minimize the aberration by scanning the skull and calculating the required profile of SOAP for phase compensation.

In summary, OFUS offers ultrahigh precision non-invasively towards neurological research in sub-regions of a brain. Its flexibility in fabrication, high spatiotemporal resolution, and improved electromagnetic compatibility further enable clinical applications, such as ultrasound surgery, drug delivery, and pain management. This work thus underlines the potential for OFUS to be utilized as a valuable technology in neuroscience research and clinical therapies.

**Materials and Methods**

**Simulation of the ultrasound field generated by SOAP**

The ultrasound field generated by SOAP was simulated in 2D by an open-source k-Wave toolbox on MATLAB R2019b (MathWorks, MA). No light absorption was considered during the simulation. In a 2D simulation, a block of PDMS with a curvature was placed at the interface of water and air. Water and air are serving as propagation medium and backing material, respectively. The density and acoustic speed of different materials were defined accordingly. The central frequency and bandwidth were set to 15 MHz and 200%, respectively.

**Fabrication of HSM-SOAP**

To form a curvature, a ~20 mm piece of heat-shrink tubing (McMaster-Carr, 6363K214) was filled with a steel bead with a diameter of 12.7 mm (McMaster-Carr, 9529K22), and heated up with a heat gun to fully shrink. After that, we cut the tubing 2 mm away from a great circle of the steel bead and obtained an HSM-SOAP.

**Fabrication of CNT-PDMS and CNP-PDMS SOAPs**

5 wt% multi-wall carbon nanotubes (VWR, MFCD06202029) and CNP (Sigma-Aldrich, 633100-25G) were mixed with the PDMS base and curing agent matrix (10:1 weight ratio, Dow Corning Corporation, Sylgard 184), respectively. The mixture was poured into a 3D-printed mold designed with a 12.7 mm diameter and 2 mm working distance and degassed in vacuum for 30 min. To get fully cured CNT-PDMS and CNP-PDMS SOAPs, the mixture was heated in an oven to 60 °C for 2.5 hrs before being removed from the mold.

**Fabrication of CS-PDMS SOAP**

A steel bead with a diameter of 12.7 mm was placed at the flame core of a paraffin wax candle for 10 to 15 s to be fully coated with flame synthesized candle soot nanoparticles. Then, the coated bead was dipped into the degassed PDMS base and curing agent matrix (10:1 weight ratio) and positioned so that the surface of the PDMS matrix is 2 mm lower than the great circle plane of the bead. The cured sample was obtained after 15 min heating at 110 °C on a heat plate.

**OFUS generation and characterization**

SOAP placed in a water tank was illuminated from the bottom. A Q-switched Nd: YAG laser (Quantel Laser, CFR ICE450) delivered 8 ns pulses at 1064 nm to SOAP to generate an optoacoustic signal. The laser was modulated by a function generator (33220A, Agilent) at a repetition rate of 10 Hz. A system consisting of a needle hydrophone, a submersible preamplifier, and a DC coupler was used for the ultrasound pressure and waveform measurement. The waveform was amplified with an ultrasonic pulser-receiver (Olympus, Model 5073PR) and collected with a digital oscilloscope after 4 times average (Rigol, DS4024). A 40 μm needle hydrophone (Precision Acoustic, NH0040, optimized for 5 to 40 MHz range) was used to acquire the waveform and pressure of SOAPs made of four materials, the transcranial efficiency, the spatial profile of the focus generated by SOAP, and the ultrasound pressure of 20 MHz transducer for neural stimulation. The needle hydrophone has an upper limit at 50 MPa of linear range. This pressure level corresponds to a laser energy of 0.65 mJ/cm$^2$, which is far below the energy range we used for later experiments. To measure the pressure generated by SOAP with the laser energy up to 8 mJ/cm$^2$, another 75 μm needle hydrophone (Precision Acoustic, NH0075, optimized for 5 to 40 MHz range) was used.

**SEM imaging of SOAP**

Before SEM imaging (Zeiss, Supra 55VP), thin cross-section slices of SOAP were sputtered with Au/Pd for 10 s and mounted on an aluminum stub. SEM images of SOAP were obtained at 3kV as accelerating voltage with an aperture size of 20 microns.

**SRS and photothermal imaging**

An 80-MHz femtosecond pulsed laser (Spectra-Physics, InSight X3) provides a tunable beam (from 680 nm to 1300 nm) and a synchronized beam (fixed at 1045 nm) for the multimodal imaging system. To image the PDMS and candle soot, the tunable beam was set to 801 nm as the pump beam, along with the fixed wavelength beam as the Stokes for femtosecond stimulated Raman scattering (SRS) imaging of C-H bonds and also the probe beam for pump-probe imaging of candle soot simultaneously. After the Stokes/probe beam was modulated by an acoustic-optic modulator (Isomet Corporation, 1205c), the pump and Stokes/probe beams were combined by a dichroic mirror and directed into a lab-built laser scanning microscope. The temporal delay between the pump and Stokes/probe pulses was controlled by a motorized delay stage. A 60× water objective (Olympus, UPlanApo 60XW, NA=1.2) focused the collinear beams onto the sample. The power of each beam on the sample was 2 mW. The two beams were collected in the forward direction by an oil condenser (Olympus, Aplanat Achromat 1.4, NA=1.4) and then filtered by short pass filters. After filtering, only the pump beam was detected by a photodiode with a laboratory-built resonant amplifier. A lock-in amplifier (Zurich Instrument, MFLI) demodulated the detected pump beam for the stimulated Raman loss signal and the pump-probe signal according to the modulation transfer. The femtosecond SRS of C-H bonds in PDMS only occurs when the two beams temporally overlapped (t ~ 0 s) while the candle soot's pump probe photothermal signal has a much longer decay. Thus, an x-y-t image stack was acquired to decompose the distribution of the CS-PDMS mixture. A chemical composition map can be generated by applying least square fitting with the time domain references from the pure samples. This strategy allowed simultaneous SRS imaging of PDMS and photothermal imaging of candle soot. This method was firstly validated with pure CS sample and CS-PDMS mixture on a glass slide. Then the cross-section of SOAP was tested.

**Optoacoustic tomography imaging**

An optoacoustic tomography system consists of a customized 128-element transducer array (L22-14v, Verasonics Inc., 50 % bandwidth) and a 128-channel ultrasound data acquisition system (Vantage 128, Verasonics Inc.). The ultrasound image was measured using the US mode, sending acoustic pulses and receiving echoes back. The OA image was taken using the receiving-only mode for OA signals acquisition. The OAT system is synchronized to the Quantel laser by a function generator and a delay generator (DG535, Stanford Research Systems). The function generator triggered the Quantel laser and the delay generator with a pulse mode at a 10 Hz repetition rate. The delay generator added another tunable delay to the Vantage 128 to receive ultrasound signals at different time delays after the optoacoustic signal was generated. By tuning the delay, propagation of the optoacoustic signal can be visualized.

**Embryonic neuron culture**

All experimental procedures have complied with all relevant guidelines and ethical regulations for animal resting and research established and approved by the Institutional animal care and use committee of Boston University (PROTO201800534). 35 mm glass-bottomed dishes were coated with 50 µg/ml poly-D-lysine (Sigma-Aldrich), placed in an incubator at 37 °C with 5% $CO_2$ overnight, and washed with sterile $H_2O$ three times before seeding the neuron. Primary cortical neurons were derived from Sprague-Dawley rats on embryonic day 18 (E18) of either sex and digested in papain (Thermo Fisher.). A medium with 10% heat-inactivated fetal bovine serum (FBS, Atlanta Biologicals), 5% heat-inactivated horse serum (HS, Atlanta Biological), 2 mM Glutamine Dulbecco's Modified Eagle Medium (DMEM, Thermo Fisher Scientific Inc.) was used for washing and triturating dissociated cells. Cells were cultured in cell culture dishes (100 mm diameter) for 30 min at 37 °C in a humid incubator to eliminate fibroblasts and glial cells. The supernatant containing neurons was collected and seeded in poly-D-lysine coated dishes with 10% FBS + 5% HS + 2 mM glutamine DMEM medium. After 16 h, the medium was replaced with Neurobasal medium (Thermo Fisher) containing 2% B27 (Thermo Fisher), 1% N2(Thermo Fisher), and 2 mM glutamine (Thermo Fisher). 5 µM 5-fluoro-2′-deoxyuridine (Sigma-Aldrich) and AAV9.Syn.Flex.GCaMP6f.WPRE.SV40 virus (Addgene, MA, USA) at 1 µl/ml final concentration was added to the medium at day 5, for preventing glial proliferation and expressing GCaMP6f, respectively. 50% of the medium was replaced with a fresh culture medium every 3 to 4 days, and neurons were used for stimulation after 10 to 13 days.

**OFUS stimulation *in vitro***

SOAP was mounted on a 3D-printed holder to a translation stage for fine adjustment of the device position. The Quantel laser was delivered to SOAP in free space for optoacoustic signal generation. An inverted microscope (Eclipse TE2000-U, Nikon) with 10X objective (Plan Fluor, 0.3 NA, 16 mm WD, Thorlabs) was used for fluorescence imaging. The microscope was illuminated by a 470-nm LED (M470L2, Thorlabs), filtered by a filter set for green fluorescent protein (MDF-GFP, Thorlabs), and imaged with a CMOS camera (Zyla 5.5, Andor). Before stimulation, the focus of ultrasound was visualized by the motion of 9.9 µm green fluorescent beads (G1000, Duke Scientific Corp) dispersed in water. The focus was then adjusted to the center of the field of view, so that once the neuron culture dish was placed on the sample stage, the neurons at the center of the view field would be at the OFUS focus. After we switched to neuron culture, SOAP was placed 2 mm above the neuron in the Z direction to ensure the focus can reach neurons. For each trial of stimulation, we delivered a single cycle of optoacoustic wave to perturb the neuron cells. The CMOS camera was synchronized with the laser. The fluorescence intensities in imaging sequences were analyzed with ImageJ (Fiji) after experiments.

**Cell viability study**

For cell viability studies, we randomly selected five groups of neurons at the OFUS focus in each dish at selected energy level. Each group of neurons was delivered 30 cycles of OFUS in total. Every 3 consecutive cycles were delivered at 10 Hz with an interval of 5 s. All the cells were labelled with GCaMP6f. The dead cells were stained with 1 µL 100 µg/Ml propidium iodide (P1304MP, Thermo Fisher Scientific Inc.) solution for 15 min. The cells were incubated for 30 min before imaged with fluorescence microscope for analysis. The live cells were distinguished with positive GCaMP6f without PI staining. The dead cells were labelled with both GCaMP6f and PI. The live and dead cells within an area of $200 \times 200$ µm$^2$ at the focus were counted.

**OFUS stimulation *in vivo***

Adult C57BL/6J mice (age 14-16 weeks) were initially anesthetized with 5 % isoflurane in oxygen and then fixed on a standard stereotaxic frame with 1.5 ~ 2 % isoflurane. A tail pinch was used to determine the anesthesia depth. A heating pad was placed under the mouse to maintain the body temperature. The hair on the targeted brain was removed. SOAP mounted on a 3D-printed holder was aligned to the targeted motor cortex of the mouse. The Quantel laser was delivered to SOAP in free space at a 10 Hz repetition rate. For c-fos expression, a pulse train was delivered with 33% duty cycle (2 s laser on, 4 s laser off) for 30 min. For EMG recording, a pulse train of 2 seconds was delivered, which includes 20 cycles of optoacoustic wave.

**EMG recording and signal processing**

The OFUS focus was aligned with the primary motor cortex (ML: 1.5, AP: 0.5). To record the muscle activity, the needle electrode was inserted subcutaneously into the hind limb biceps femoris muscle and the ground electrode was inserted into the tail. The control group was recorded on the trunk ipsilateral to the stimulation site. EMG signals were recorded by a Multi-Clamp 700B amplifier (Molecular Devices), filtered at 1 to 5000 Hz, digitized with an Axon DigiData 1,550 digitizer (Molecular Device), and filtered by a noise eliminator (D-400, Digitimer). EMG signal was filtered with bandpass filter at 0.5 ~ 500 Hz and full-wave rectification. Then the envelope of the processed signal was plotted.

**Immunofluorescence staining and imaging**

After the stimulation session, the mouse was put to rest for 1 h for maximized c-fos expression, then was sacrificed and perfused transcardially with phosphate-buffered saline (PBS, 1X, PH 7.4, Thermo Fisher Scientific Inc.) solution and 10% formalin. After fixation, the brain was extracted and fixed in 10% formalin solution for 24 hrs. The fixed mouse brain was immersed in 1X PBS solution. The brain was sliced to coronal sections with a 150 μm thickness using an Oscillating Tissue Slicer (OST-4500, Electron Microscopy Sciences). Brain slices were gently transferred by a brush into 10% formalin solution for another 24 h fixation, then blocked with 5% Bovine serum albumin (Sigma-Aldrich)-PBS solution for 30 min at room temperature. The slices were permeabilized with 0.2% Triton (Triton X-100, 1610407, Bio-Rad Laboratories)-PBS solution for 10 min, and incubated with Anti c-Fos Rabbit antibody (4384S, Cell Signaling Technology) at a concentration of 2 μg/mL at 4 °C overnight. Following primary incubation, slices were incubated with secondary antibody Alexa Fluor 488 goat anti-rabbit IgG (Thermo Fisher Scientific) at a concentration of 1 μg/mL and DAPI (Thermo Fisher Scientific) at 5 μg/mL in dark for 2 h at room temperature. Between steps, the slices were rinsed with 0.2% Tween (Tween 20, Tokyo Chemical Industry)-PBS solution 4 times for 5 min. Fluorescent images were acquired with an FV3000 Confocal Laser Scanning Microscope (Olympus). Confocal images were acquired under an excitation wavelength of 405 nm for DAPI and 488 nm for c-Fos.

**Histology examination**

After the stimulation session, the mouse was sacrificed immediately, and perfused and fixed as previously described. The brain was embedded in paraffin and sliced for 5 μm thickness at 150 μm step to obtain coronal sections. Slicing and standard H&E staining were performed at Mass General Brigham Histopathology Research Core. Histology images were acquired with a VS120 Automated Slide Scanner (Olympus).

**Temperature measurement**

A thermocouple (DI-245, DataQ) was used to record the temperature profile. The tip of the thermocouple was placed at the focus and the surface of SOAP, respectively. The pressure of 52.7 MPa was delivered for 10 s at 10 Hz during recording. The temperature at the SOAP surface without light was also recorded as a baseline.

**Statistical analysis**

Acoustic waveforms, calcium traces, and electrophysiological traces were plotted with Origin 2020. FFT for the frequency spectrum was performed with MATLAB R2019b. Data are presented with the mean ± standard error of the mean. Calcium images were processed with ImageJ. All comparative data were analyzed with a two-sample $t$-test. $p$ values were defined as: n.s., not significant ($p > 0.05$); * $p < 0.05$; ** $p < 0.01$; *** $p < 0.001$.

**References**

1. Mehić, E. *et al*. Increased anatomical specificity of neuromodulation via modulated focused ultrasound. *PLoS One* **9**, e86939 (2014).
2. Choi, T. *et al*. A soft housing needle ultrasonic transducer for focal stimulation to small animal brain. *Annals of Biomedical Engineering* **48**, 1157-1168 (2020).
3. Arcot Desai, S. *et al*. Deep brain stimulation macroelectrodes compared to multiple microelectrodes in rat hippocampus. *Frontiers in Neuroengineering* **7**, 16 (2014).
4. Leuthardt, E. C. *et al*. Evolution of brain-computer interfaces: going beyond classic motor physiology. *Neurosurgical Focus* **27**, E4 (2009).
5. Rosin, B. *et al*. Closed-loop deep brain stimulation is superior in ameliorating parkinsonism. *Neuron* **72**, 370-384 (2011).
6. Boon, P. *et al*. Deep brain stimulation in patients with refractory temporal lobe epilepsy. *Epilepsia* **48**, 1551-1560 (2007).
7. Ineichen, C., Shepherd, N. R. & Sürücü, O. Understanding the effects and adverse reactions of deep brain stimulation: is it time for a paradigm shift toward a focus on heterogenous biophysical tissue properties instead of electrode design only? *Frontiers in Human Neuroscience* **12**, 468 (2018).
8. Boyden, E. S. *et al*. Millisecond-timescale, genetically targeted optical control of neural activity. *Nature Neuroscience* **8**, 1263-1268 (2005).
9. Chen, R. *et al*. Deep brain optogenetics without intracranial surgery. *Nature Biotechnology* **39**, 161-164 (2021).
10. Hallett, M. Transcranial magnetic stimulation and the human brain. *Nature* **406**, 147-150 (2000).
11. Tufail, Y. *et al*. Transcranial pulsed ultrasound stimulates intact brain circuits. *Neuron* **66**, 681-694 (2010).
12. Tufail, Y. *et al*. Ultrasonic neuromodulation by brain stimulation with transcranial ultrasound. *Nature Protocols* **6**, 1453-1470 (2011).
13. Yoo, S. S. *et al*. Non-invasive brain-to-brain interface (BBI): establishing functional links between two brains. *PLoS One* **8**, e60410 (2013).
14. Yoo, S. S. *et al*. Focused ultrasound modulates region-specific brain activity. *Neuroimage* **56**, 1267-1275 (2011).
15. Deffieux, T. *et al*. Low-intensity focused ultrasound modulates monkey visuomotor behavior. *Current Biology* **23**, 2430-2433 (2013).
16. Legon, W. *et al*. Transcranial focused ultrasound modulates the activity of primary somatosensory cortex in humans. *Nature Neuroscience* **17**, 322-329 (2014).
17. Lee, W. *et al*. Simultaneous acoustic stimulation of human primary and secondary somatosensory cortices using transcranial focused ultrasound. *BMC Neuroscience* **17**, 68 (2016).
18. Rezayat, E. & Toostani, I. G. A review on brain stimulation using low intensity focused ultrasound. *Basic and Clinical Neuroscience* **7**, 187-194 (2016).
19. Li, G. F. *et al*. Improved anatomical specificity of non-invasive neuro-stimulation by high frequency (5 MHz) ultrasound. *Scientific Reports* **6**, 24738 (2016).
20. Menz, M. D. *et al*. Radiation force as a physical mechanism for ultrasonic neurostimulation of the *ex vivo* retina. *Journal of Neuroscience* **39**, 6251-6264 (2019).
16

**Acknowledgments**

**Funding:** This work is supported by R01 NS109794 to J.-X.C. and C. Y. and R01 HL125385 to J.-X.C. Research reported in this publication was supported by the Boston University Micro and Nano Imaging Facility and the Office of the Director, National Institutes of Health of the National Institutes of Health under award Number S10OD024993. The content is solely the responsibility of the authors and does not necessarily represent the official views of the National Institute of Health.

**Author contributions:** LL and JXC conceived the idea of SOAP; YL simulated, fabricated, and characterized SOAP; YL and YJ designed and performed biology experiments; XG performed SRS and photothermal experiments; YL and RC performed SEM imaging; YL and YZ performed confocal imaging; YL, YJ, LL, XG, YZ, NZ, CY and JXC discussed and analyzed the data; YL and LS prepared neuron cultures; GC and RW helped with data collection; YL wrote the original draft of manuscript; LL, YJ, GC, CY and JXC reviewed and edited the manuscript. CY and JXC supervised the project. We appreciate Carolyn Marar for language editing.

**Competing interests:** Authors declare that they have no competing interests.

**Data and materials availability:** All data are available in the main text or the supplementary materials.

**Figures and Tables**

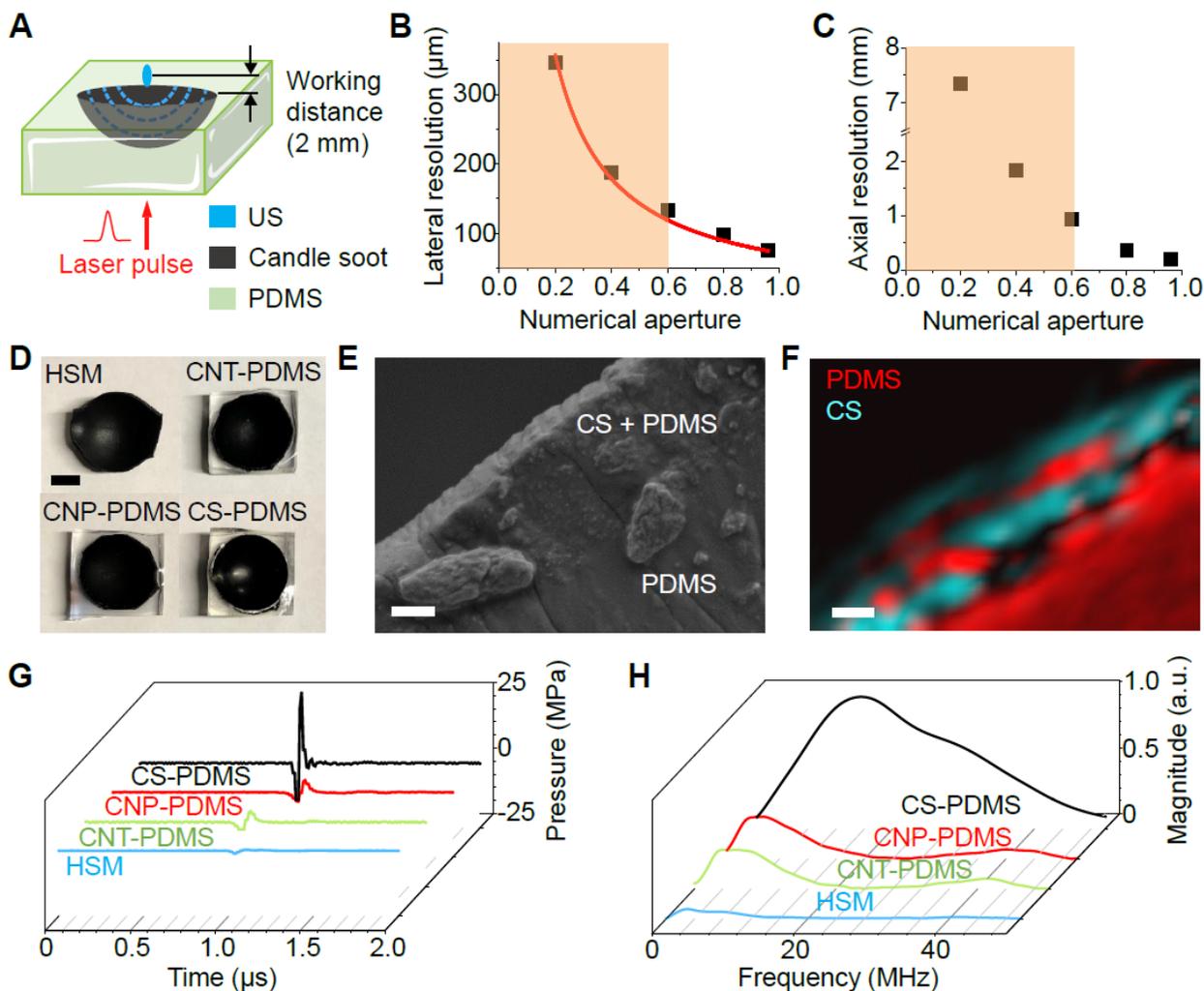

**Figure 1. Design, fabrication, and characterization of SOAP.** (**A**) The schematic of SOAP design. (**B**) Numerical aperture and lateral resolutions. Red line: fitting curve. Orange area: the NA range of conventional ultrasound transducers. (**C**) Numerical aperture and axial resolutions. (**D**) Photos of 4 kinds of SOAPs with the same geometric design. From left to right, top to bottom: heat shrink membrane (HSM), carbon nanotube-PDMS (CNT-PDMS), carbon nanoparticles-PDMS (CNP-PDMS), candle soot-PDMS (CS-PDMS). Scale bar: 5 mm. (**E**) SEM image of the CS-PDMS SOAP cross-section. Scale bar: 1 μm. (**F**) The spatial distribution of PDMS and CS in SOAP cross-section by SRS and photothermal imaging. Red: PDMS. Cyan: CS. Scale bar: 1 μm. (**G**) The ultrasound waveforms from 4 kinds of SOAPs with the same laser energy input. (**H**) The frequency spectra of ultrasound from 4 kinds of SOAPs.

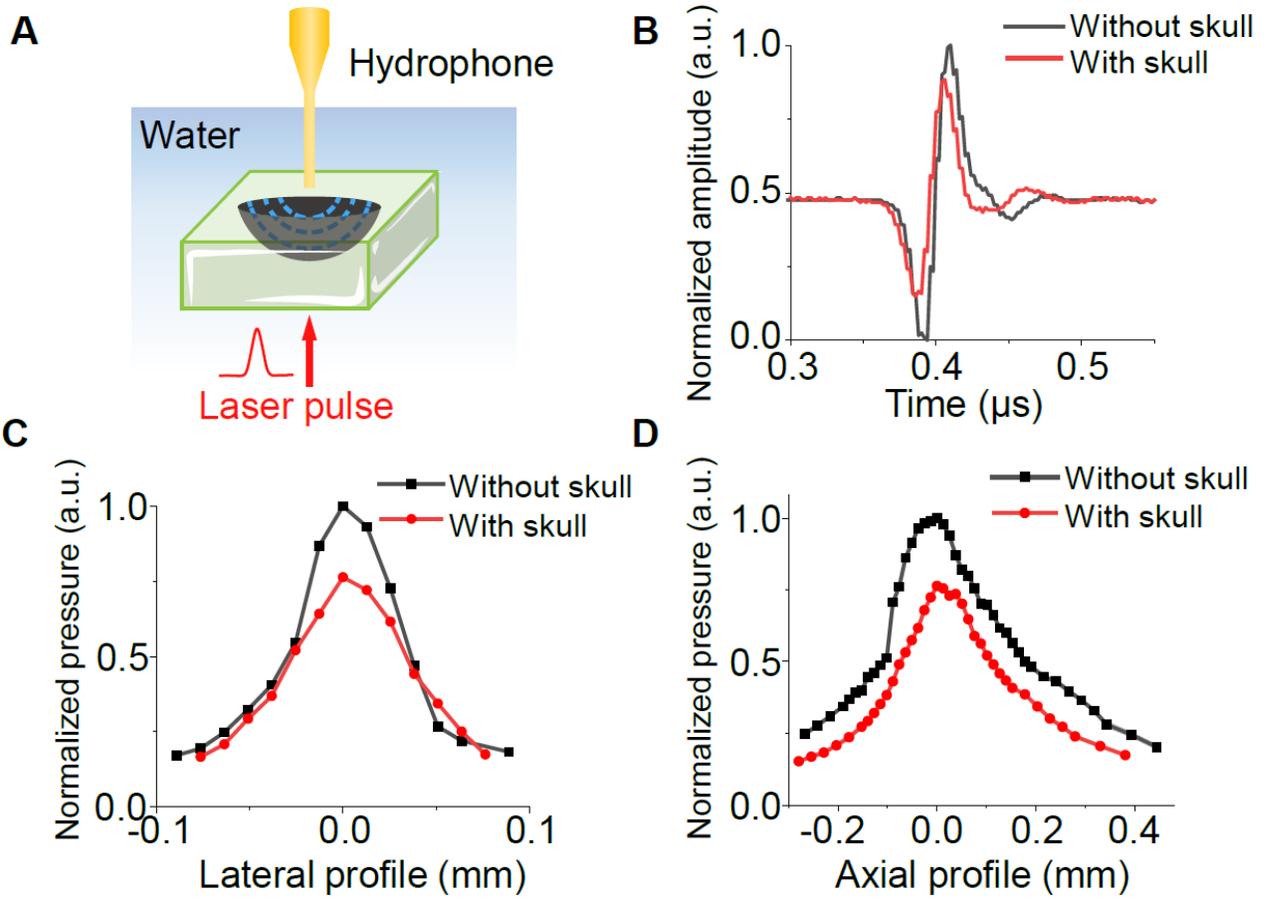

**Figure 2. Transcranial efficiency and high spatial resolution of OFUS penetrating a mouse skull.** (**A**) The illustration of the experimental setup for characterizing OFUS with a needle hydrophone. (**B**) Time domain optoacoustic signals were measured at the focus without and with a piece of mouse skull. Results were normalized to the peak pressure without the skull. (**C**) The lateral resolution of OFUS without and with a piece of mouse skull. (**D**) The axial resolution of OFUS without and with a piece of mouse skull.

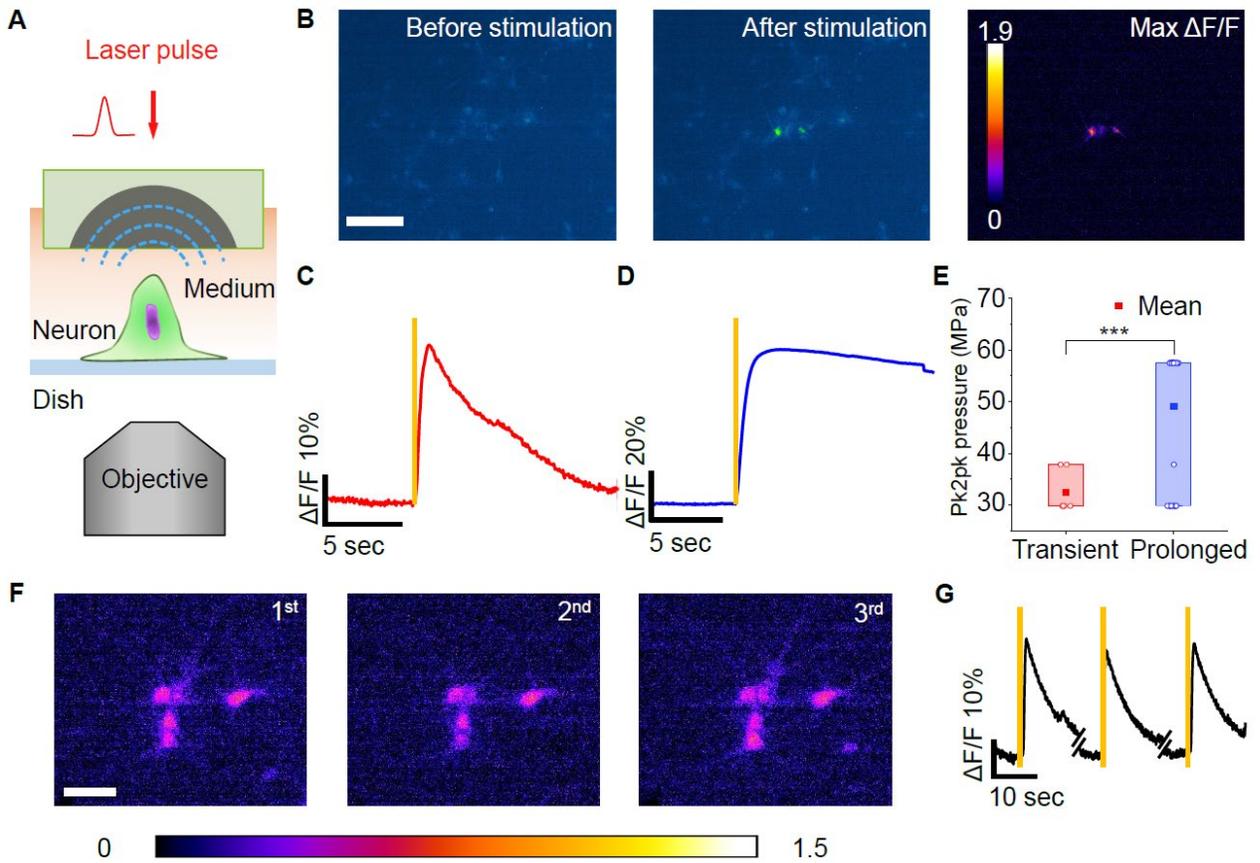

**Figure 3. Cultured neurons are stimulated by OFUS delivered by SOAP *in vitro*.** (**A**) The schematic of OFUS stimulation setup. (**B**) Representative calcium images of neurons before and after OFUS stimulation. The right panel shows the maximum ΔF/F. Scale bar: 50 μm. (**C**) The average calcium trace of transient stimulation. (**D**) The average calcium trace of prolonged OFUS stimulation. Average traces in solid and SEM in shades. Yellow vertical lines: laser on. (**E**) Statistics of the threshold pressure of transient and prolonged stimulation. ***$p < 0.001$. (**F**) Maximum ΔF/F image of repeated stimulation for the safety demonstration. Scale bar: 50 μm. (**G**) Calcium traces corresponding to the repeated stimulation in E were taken from the same neuron. Yellow vertical lines: laser on.

| Device | Frequency (MHz) | $I_{SPPA}$ (W/cm$^2$) | Duration (s) | Total energy density (J/cm$^2$) |
|---|---|---|---|---|
| SOAP (this work) | 15 | $6.3 \times 10^3$ | $0.09 \times 10^{-6}$ | $5.7 \times 10^{-4}$ |
| Transducer (this work) | 20 | 32 | 0.5 | 16 |
| Transducer [29] | 0.3 | 15 | 0.5 | 7.5 |

**Table 1. Experimental conditions and the total energy density to evoke a similar amplitude of neuron responses.** The total energy density equals to the ultrasound intensity times the duration.

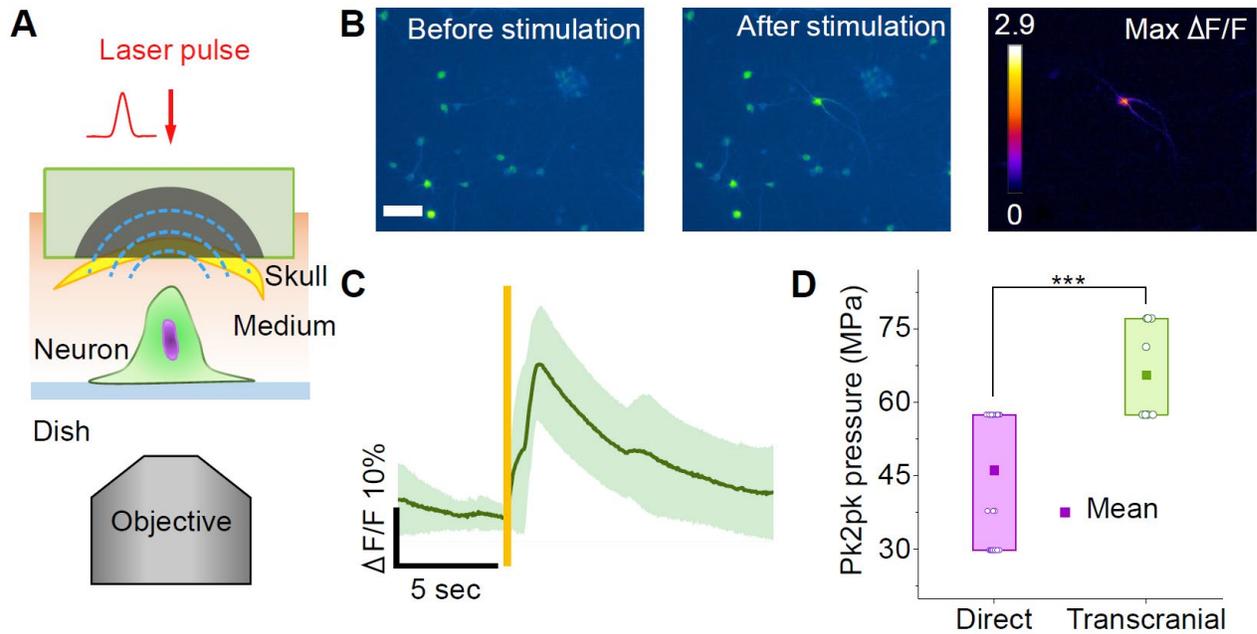

**Figure 4. Transcranial stimulation by OFUS *in vitro*.** (**A**) The schematic of transcranial *in vitro* stimulation. (**B**) Representative images of neurons before and after transcranial stimulation. The rightest panel showed the maximum ΔF/F. Scale bar: 50 μm. (**C**) The averaged calcium response trace of transcranial OFUS stimulation. Yellow vertical line: laser on. The average trace in solid and SEM in shades. (**D**) Statistics of threshold pressure of direct and transcranial stimulation with a single cycle. ***$p < 0.001$.

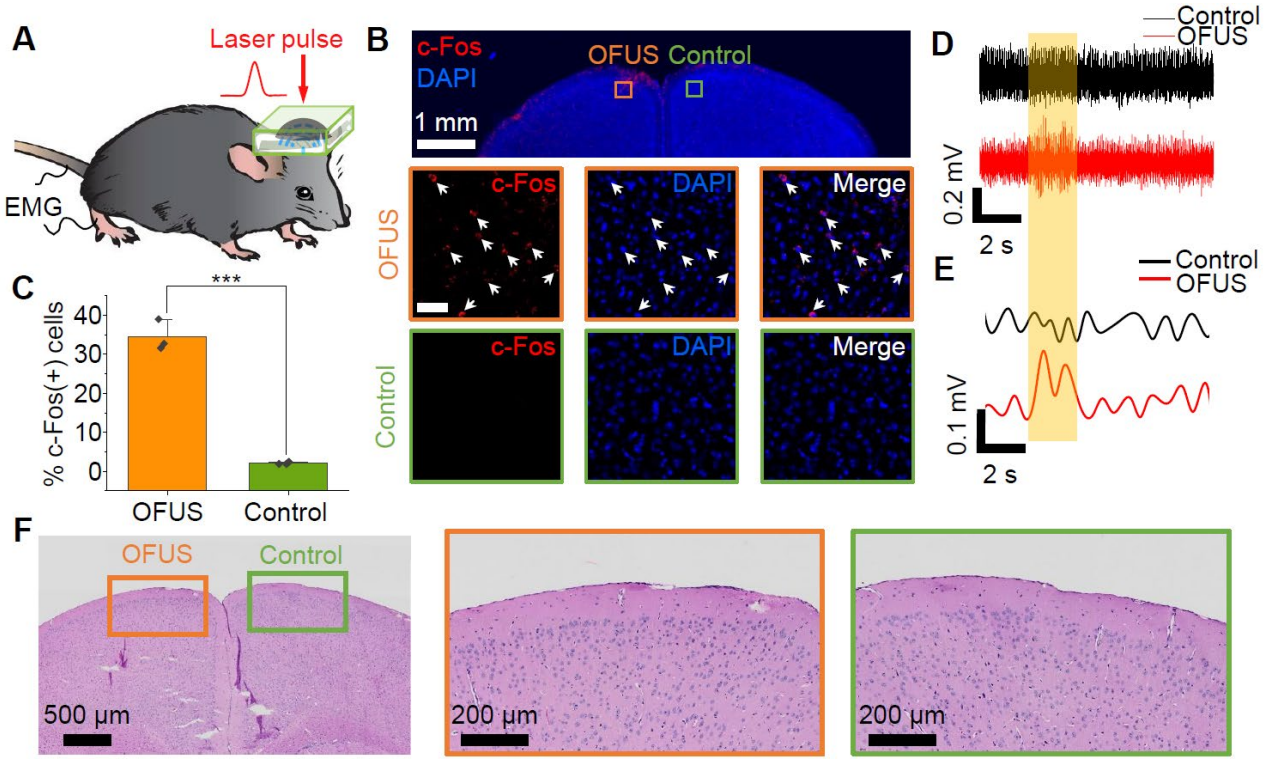

**Figure 5. Representative immunofluorescence examinations and EMG recording trace for *in vivo* OFUS stimulation.** (**A**) The schematic of OFUS stimulation *in vivo*. (**B**) Representative images of c-Fos and DAPI staining within the stimulation and control area. Red: c-Fos. Blue: DAPI. Orange outline: OFUS stimulated area. Green outline: control group at the contralateral area. Scale bar: middle and lower panel, 50 μm. (**C**) Statistic analysis of the percentage of c-Fos positive neurons. ***p<0.001, two-sample *t*-test. (**D**) Representative EMG recordings of 2 s OFUS stimulation and control group targeting the somatosensory cortex. Orange box: laser on. (**E**) EMG signals after the band-pass filter and full-wave rectifier and envelope. (**F**) Histology results after stimulation *in vivo*.